\documentclass[12pt,preprint]{aastex}

\usepackage[dvipdfm,bookmarks=true,bookmarksnumbered=true,bookmarkstype=toc,colorlinks=true,linkcolor=blue,urlcolor=blue,citecolor=blue,]{hyperref}

\newcommand\mearth {{M_\oplus}}
\newcommand\ltsima{$\; \buildrel <\over\sim \;$}
\newcommand\simlt{\lower.5ex\hbox{\ltsima}}
\newcommand\gtsima{$\; \buildrel >\over\sim \;$}
\newcommand\simgt{\lower.5ex\hbox{\gtsima}}

\begin{document}

\title{A sub-Saturn Mass Planet, MOA-2009-BLG-319Lb}

\author{ 
N.~Miyake\altaffilmark{1,72}, 
T.~Sumi\altaffilmark{1,72}, 
Subo~Dong\altaffilmark{2,3,73},
R.~Street\altaffilmark{4,5,74},  
L.~Mancini\altaffilmark{6,7,8,75}, 
A.~Gould\altaffilmark{9,73}, 
D.~P.~Bennett\altaffilmark{10,72,76}, 
Y.~Tsapras\altaffilmark{4,11,74},  
J.~C.~Yee\altaffilmark{9,73}, 
M.~D.~Albrow\altaffilmark{12,76}
% ...
\\
}

\altaffiltext{1}{Solar-Terrestrial Environment Laboratory, Nagoya University, Nagoya 464-8601, Japan; nmiyake,sumi,furusawa,dsuke,abe,afukui,fhayashi,hosaka,itow,kkamiya,makita,kmasuda,ymatsu,nishimo@stelab.nagoya-u.ac.jp}
\altaffiltext{2}{Institute for Advanced Study, Einstein Drive, Princeton, NJ 08540, USA; dong@ias.edu}
\altaffiltext{3}{Sagan Fellow}
\altaffiltext{4}{Las Cumbres Observatory Global Telescope Network, 6740 Cortona Dr., Suite 102, Goleta, CA 93117}
\altaffiltext{5}{Dept. of Physics, Broida Hall, University of California, Santa Barbara CA 93106-9530, USA}
\altaffiltext{6}{Universit\`{a} degli Studi di Salerno, Dipartimento di Fisica "E.R.~ Caianiello", Via Ponte Don Melillo, 84085 Fisciano (SA), Italy}
\altaffiltext{7}{Istituto Internazionale per gli Alti Studi Scientifici (IIASS), Via G.\ Pellegrino 19, 84019 Vietri sul Mare (SA), Italy}
\altaffiltext{8}{Dipartimento di Ingegneria, Universit\`{a} del Sannio, Corso Garibaldi 107, 82100 Benevento, Italy}
\altaffiltext{9}{Department of Astronomy, Ohio State University, 140 W.\ 18th Ave., Columbus, OH 43210, USA; gould,gaudi,jyee,pogge,simkoz@astronomy.ohio-state.edu}
\altaffiltext{10}{Department of Physics, University of Notre Dame, Notre Dame, IN 46556, USA; bennett@nd.edu}
\altaffiltext{11}{Astronomy Unit, School of Mathematical Sciences, Queen Mary, University of London, London E1 4NS}
\altaffiltext{12}{Department of Physics and Astronomy, University of Canterbury, Private Bag 4800, Christchurch, New Zealand}
% ...
\altaffiltext{72}{Microlensing Observations in Astrophysics (MOA) Collaboration, \url{http://www.phys.canterbury.ac.nz/moa}}
\altaffiltext{73}{Microlensing Follow Up Network ($\mu$FUN), \url{http://www.astronomy.ohio-state.edu/\~microfun}}
\altaffiltext{74}{RoboNet, \url{http://robonet.lcogt.net}}
\altaffiltext{75}{Microlensing Network for the Detection of Small Terrestrial Exoplanets (MiNDSTEp), \url{http://www.mindstep-science.org}}
\altaffiltext{76}{Probing Lensing Anomalies Network (PLANET), \url{http://planet.iap.fr}}

\begin{abstract}
We report the gravitational microlensing
discovery of a sub-Saturn mass planet, MOA-2009-BLG-319Lb, 
orbiting a K or M-dwarf star in the inner Galactic disk or Galactic bulge.
The high cadence observations of the MOA-II survey discovered this
microlensing event and enabled its identification as a high magnification
event approximately 24 hours prior to peak magnification. As a result, 
the planetary signal at the peak of this
light curve was observed by 20 different telescopes, which is 
the largest number of telescopes to contribute to a planetary discovery 
to date. The microlensing model for this event indicates a planet-star 
mass ratio of $q = (3.95 \pm 0.02) \times 10^{-4}$ and a separation of 
$d = 0.97537 \pm 0.00007$ in units of the Einstein radius. A Bayesian analysis 
based on the measured Einstein radius crossing time, $t_{\rm E}$, 
and angular Einstein radius, $\theta_{\rm E}$, along with a standard
Galactic model indicates a host star mass of 
$M_{\rm L} = 0.38_{-0.18}^{+0.34}$ $M_\odot$ and a planet mass of
$M_{\rm p} = 50_{-24}^{+44}$ $M_\oplus$, which is half the mass of 
Saturn. This analysis also yields a planet-star
three-dimensional separation of $a = 2.4_{-0.6}^{+1.2}$ AU and
a distance to the planetary system of $D_{\rm L} = 6.1_{-1.2}^{+1.1}$ kpc. 
This separation is $\sim 2$ times the distance of the snow line, a separation similar
to most of the other planets discovered by microlensing.
\end{abstract}


\section{Introduction}
We present the eleventh microlensing planet, following ten previous discoveries
\citep{Bond2004,ogle71,ogle390,Gould2006,Gaudi2008,
Bennett2008,dong-moa400,Sumi2010,moa310}.
Microlensing is unique among exoplanet detection methods in that it is
sensitive to planets with masses down to $1~M_\oplus$ 
\citep{Bennett1996} at relatively large separations, typically
between 1~AU and 6~AU, depending on the mass of the host star. 
These separations are generally beyond the 
``snow line" at $\sim 2.7$~AU~$M/M_\odot$
\citep{ida_lin,lecar_snowline,kennedy_snowline}, the region where planets
can form most quickly, according to the core accretion theory.
Microlensing confirms this expectation, as a statistical analysis of the
prevalence of planets found by microlensing shows that Saturn-mass
planets beyond the snow line are more common than the higher mass
gas giants found by radial velocities in shorter period orbits
\citep{Gould2010}, although the microlensing results are consistent
with an extrapolation of the radial velocity results for solar-mass stars 
to larger orbital distances \citep{Cumming2008}.
Furthermore, \citet{Sumi2010} have shown
that the number of planets (per logarithmic interval)
increases with decreasing mass ratio, $q$,
as $q^{-0.7\pm 0.2}$, down to $\sim 10~\mearth$. So, cold Neptunes
seem to be even more common than cold Saturns. While the
number of planets found by microlensing is relatively small, it is
the cold-Neptunes and Saturns discovered by microlensing that
represent the most common types of exoplanet yet to be discovered. 
Microlensing has also found the first Jupiter/Saturn analog planetary
system  \citep{Gaudi2008,bennett-ogle109}, and it should soon be
possible to use the microlensing results to determine how the properties
of exoplanet systems vary with distance from the Galactic center.

Searches for exoplanets via the microlensing method are currently conducted by
two survey groups, the Microlensing Observations in Astrophysics 
(MOA; \citealt{Bond2001,Sumi2003}) and the Optical Gravitational Lensing 
Experiment (OGLE; \citealt{Udalski2003}), which monitor 
$\sim 40\,{\rm deg}^2$ of the Galactic bulge to identify stellar microlensing
events that can be searched for planetary signals. The planetary signals
have durations that range from a few hours to a few days, so a global
network of telescopes is needed to search for and characterize planetary
signals. The follow-up groups that complete this telescope network are the
Microlensing Follow-Up Network ($\mu$FUN), 
RoboNet, Microlensing Network for the Detection of Small Terrestrial 
Exoplanets (MiNDSTEp), and the Probing Lensing Anomalies NETwork (PLANET). These narrow field-of-view 
follow-up telescopes can provide very high cadence observations of a small number
of events that are known to be interesting, due to known or suspected
planetary deviations in progress \citep{Sumi2010} or high magnification
events, which have very high planet detection efficiency
\citep{griest_saf,mps-98blg35,Rattenbury2002}. The very wide ($2.2\,{\rm deg}^2$)
field-of-view of the MOA-II 1.8m telescope with
80M pixel CCD camera MOA-cam3 \citep{Sako2008} provides
high cadence survey observations of the entire Galactic bulge, and this
allows MOA to identify suspected planetary deviations in progress and 
to predict high magnification ($A_{\rm max} \simgt 100$) for events
with relatively short timescales (Einstein radius crossing time $t_{\rm E} < 20\,$days.). 
MOA-2009-BLG-319 is one such short timescale high magnification event
that was identified as a high magnification event based on MOA data taken
$\sim 24$ hours prior to peak magnification.

In this paper, we report the discovery of a sub-Saturn mass planet in 
the microlensing event, MOA-2009-BLG-319. We describe the observations 
and data sets in Section \ref{sec:observation}. The light curve modeling is 
presented in Section \ref{sec:modeling}. We discuss the measurement of 
the source magnitude and color in Section \ref{sec:correction}, 
and derive the angular Einstein radius in Section \ref{sec:thetaE}. 
In Section \ref{sec:parallax}, we search for a microlensing parallax signal. In 
Section \ref{sec:estimate}, we use a Bayesian analysis to estimate
the masses and distances of the host star and the planet, based on 
the angular Einstein radius and microlens parallax.
We present our conclusion in Section \ref{sec:conclusion}.

\section{Observations}
\label{sec:observation}

For the bulk of the 2009 observing season, the MOA group was the only
microlensing survey group in operation because the OGLE group completed
the OGLE-III survey on 3 May 2009, in order to upgrade to the OGLE-IV
camera with a much wider field-of-view. 
Prompted in part by this fact, MOA adopted
a new observing strategy for the 2009 observing season in order to increase the planet detection
efficiency. The top 6 fields (a total of $13.2\,{\rm deg}^2$) yielded 54\% of the
microlensing events found by MOA in previous seasons, and these were 
observed every 15 minutes. The next 6 fields (with 25\% of the previous 
years' events) were observed every 47 minutes, and most of the remaining 
10 fields were observed every 95 minutes. This new observing strategy yielded
563 microlensing alert events in 2009, an increase of about 100 over the 2008
total. MOA-2009-BLG-319 was the first of four of these events to yield an
apparent planetary signal.

The event MOA-2009-BLG-319 
$[$($\mathrm{R.A., decl.} )_{{\rm J}2000.0}$=(18$^h$06$^m$58$^s$.13, -26$^\circ$49'10".89), 
($l$, $b$)=(4.202, -3.014)$]$ was detected 
and announced as a normal microlensing 
alert event by the MOA collaboration on 20 June 2009 
(${\rm HJD'} \equiv  {\rm HJD} - 2450000 = 5003.056$). The discovery 
announcement provided a model for this event, which 
indicated that this was a high-magnification event, and so MOA
immediately began follow-up observations 
in the $I$ and $V$-bands with the University of Canterbury's
0.6m Boller \& Chivens (B\&C) telescope
at Mt.~John Observatory. Public access to the MOA photometry over the
next two nights,
led the $\mu$FUN, RoboNet, and MiNDSTEp collaborations to begin 
observations of this event $\sim 2.5\,$days after its discovery.
Three days after the discovery, the MOA data indicated that this event was
quite likely to reach high magnification,
and the $\mu$FUN group issued a high-magnification 
alert by email to all interested observers, estimating
a peak magnification of A$_{\rm {max}} >$ 100 (at 1-$\sigma$) 18 hours later
at ${\rm HJD'} = 5006.875$. This alert message noted ``low-level systematics"
in the MOA data, which were, in fact, not systematic errors at all. Instead, this
light curve feature was the first (weak) planetary caustic crossing.
Then, 14 hours later at June 24 UT 01:12 ${\rm HJD'} \simeq 5006.55$), 
data from the $\mu$FUN SMARTS CTIO telescope in Chile provided 
clear evidence for a second, much stronger, caustic crossing feature, 
which prompted $\mu$FUN to issue an anomaly alert. This feature was readily 
confirmed by the MiNDSTEp observer at La Silla from data already in hand 
(see Fig.~\ref{fig:lightcurve}). 
A large number
of telescopes responded to this anomaly alert, resulting in continuous
photometric monitoring of the remainder of the planetary signal with
no gaps larger than 5 minutes until after the planetary signal finished,
some $\sim 20\,$hours later.

The complete data set for MOA-2009-BLG-319 consists of observations
from 20 different observatories representing 
the MOA, $\mu$FUN, RoboNet, MiNDSTEp, PLANET groups, as well as the
InfraRed Survey Facility (IRSF) telescope in South Africa. Specifically, the
data set includes data from the following telescopes and passbands: 
MOA-II (New Zealand) 1.8m wide-$R$-band, 
the Mt. John Observatory B$\&$C (New Zealand) 0.61m $I$ and $V$ bands, 
$\mu$FUN Auckland Observatory (New Zealand) 0.4m $R$-band, 
$\mu$FUN Bronberg (South Africa) 0.35m unfiltered, 
$\mu$FUN SMARTS CTIO (Chile) 1.3m $V$, $I$, and $H$ bands, 
$\mu$FUN Campo Catino Austral (CAO, Chile) 0.5m unfiltered, 
$\mu$FUN Farm Cove (New Zealand) 0.35m unfiltered, 
$\mu$FUN IAC80 (Tenerife, Spain) 0.8m $I$ band, 
$\mu$FUN Mt.Lemmon (Arizona, U.S.A.) 1.0m $I$ band, 
$\mu$FUN Southern Stars Observatory (SSO, Tahiti) 0.28m unfiltered, 
$\mu$FUN Vintage Lane Observatory (New Zealand) 0.41m unfiltered, 
$\mu$FUN Wise (Israel) 0.46m unfiltered, 
$\mu$FUN Palomar (U.S.A) 1.5m $I$ band, 
RoboNet Faulkes Telescope North (FTN, Hawaii) 2.0m SDSS-$I$ band, 
RoboNet Faulkes Telescope South (FTS, Australia) 2.0m SDSS-$I$ band, 
RoboNet Liverpool Telescope (La Palma) 2.0m SDSS-$I$ band, 
MiNDSTEp Danish (La Silla) 1.54m $I$ band, 
PLANET Canopus (Australia) 1.0m $I$ band, 
PLANET SAAO (South Africa) 1.0m $I$ band, 
and IRSF (South Africa) 1.4m $J$, $H$ and $K_{\rm S}$ bands. 
This is more follow-up telescopes than have been used for
previous planetary microlensing discoveries. 

The light curve for this event had four distinct caustic crossing features, 
which were all observed with good-to-excellent sampling. The first is
a weak caustic entry  at ${\rm HJD'} \sim 5006.05$, which is observed 
by MOA. The second is a caustic exit at magnification $A\sim 60$ 
at ${\rm HJD'}\sim5006.6$. This region of the light curve is covered by 
the CTIO, Danish, Liverpool and Wise telescopes. The next light curve
feature is a strong caustic entry, which produced the light curve peak at
$A_{\rm max}\sim 205$, at ${\rm HJD'}\sim 5006.96$. The final
caustic exit occurs shortly thereafter at ${\rm HJD'} \sim 5007.0$ at a
magnification of $A\sim 180$. This main peak covering the third and fourth
caustic crossing has excellent coverage, observed by 16 telescopes. 

The images were reduced using several different photometry methods. 
The MOA data sets were reduced by the MOA Difference Image Analysis 
(DIA) pipeline \citep{Bond2001}. The $\mu$FUN data sets 
except the CTIO $H$ band and Bronberg were reduced by the MOA DIA
pipeline and pySIS version 3.0 \citep{Albrow2009}, 
which is based on the numerical kernel method invented by 
\citet{Bramich2008}. The CTIO $H$ band and Bronberg 
data sets were reduced using the OSU DoPHOT pipeline. The Danish 
data were reduced by the DIAPL image subtraction software 
\citep{Wozniak2000}. The RoboNet and PLANET data
sets were reduced by pySIS version 3.0. The IRSF data set was reduced 
by the DoPHOT pipeline. The error bars for the data points are re-normalized 
so that $\chi^2$ per degree of freedom for the best fit model is nearly one.

All of these data sets are used for modeling
except for the CTIO $V$ and $H$ band, the Canopus and SAAO $I$ band,
and the IRSF $J$,$H$,$K_{\rm S}$ bands. The CTIO $V$-band, 
the Canopus and SAAO $I$-band, and IRSF $J$, $H$,
$K_{\rm S}$-band data sets do not have
many observations and do not cover the planetary deviation region of the
light curve. The CTIO $H$-band data was not used in the modeling
because of a cyclic pattern caused by intrapixel sensitivity variations and
image dithering. For our modeling of microlensing parallax effects, we 
have used a binned data set in order to speed up the modeling calculations. 
Note that we checked that an analysis with unbinned data gives the same results.

\section{Modeling}
\label{sec:modeling}

Inspection of Figure \ref{fig:lightcurve} indicates that the event exhibits a
number of
caustic crossings, so we expect that this event, like most planetary
microlensing events will exhibit significant finite source effects.
The first step in modeling is therefore to measure the 
source color, which then enables us to determine the limb darkening 
parameters for the various light curves.

\subsection{Source Color}
\label{sec:color}

Once a microlensing model is found, the dereddened source color and 
magnitude $[I, (V-I)]_0$ can be determined by comparing the instrumental 
values of these quantities to those of the red clump \citep{Yoo2004}. This is 
described in Section \ref{sec:correction}.  However, before a good model 
can be found, the limb-darkening coefficients must be determined, which 
requires an estimate of the source color. This seemingly endless loop can
be broken by making a model-independent measurement of the 
instrumental source color from a regression of $V$-band flux on 
$I$-band flux (and then comparing this value to the instrumental clump color). 
We find $(V-I)_0 = 0.82$, as reported in greater detail in Section \ref{sec:correction}.

\subsection{Limb Darkening}

We adopt a two-parameter square-root law \citep{Claret2000} 
for the surface brightness of the source,
\begin{equation}
S_\lambda (\vartheta ) = S_\lambda (0)\left[ 1-c(1-\cos\vartheta )-d(1-\sqrt{\cos\vartheta}) \right].
\label{eq:LD}
\end{equation}
Here, $c$ and $d$ are the limb darkening coefficients, $S_\lambda (0)$ is 
the central surface brightness of the source, and $\vartheta$ is the angle 
between the normal to the stellar surface and the line of sight, i.e., 
$\sin \vartheta$=$\theta$/$\theta_*$, where $\theta$ is the angular 
distance from the center of the source.

Based on the source color estimate of $(V-I)=0.82$, the source is likely to
have a G8 spectral type and an effective temperature of 
$T_{\rm eff} = 5475\,$K according to \citet{Bessell1988}.
We use limb darkening parameters from 
\citet{Claret2000} for a source star with effective temperature 
$T_{\rm eff} = 5500\,$K, surface gravity $\log g = 4.5$ and 
metallicity $\log[M/H] = 0.0$ as presented in Table \ref{tb:LDparameter}.

\subsection{Best Fit Model}

We search for the best fit binary lens model using a variation of the 
Markov Chain Monte Carlo (MCMC) algorithm 
\citep{Verde2003} due to \citet{DoranMueller2004} and \citet{Bennett2010} that frequently
changes the ``jump function" in order to find the $\chi^2$ minimum more quickly. 
There are three lensing parameters in common with single lens events,
the time of the closest approach to the center of mass $t_0$, 
the Einstein crossing time $t_{\rm E}$,
and the minimum impact parameter $u_0$. Binary lens models require 
four additional parameters: the planet-star mass ratio $q$, the binary lens 
separation $d$, which is projected onto the plane of the sky and normalized 
by the angular Einstein radius $\theta_{\rm E}$, the angle of the source trajectory 
relative to the binary lens axis $\alpha$, and source radius relative to the 
Einstein radius $\rho = \theta_*/\theta_{\rm E}$. In addition, for each data set and
pass band, there are two parameters to describe the unmagnified source
and background fluxes in that band.

We begin by conducting a very broad parameter search. 
The parameter search has been conducted by two independent codes.
We perform 300 separate $\chi^2$ 
minimizations with initial parameters distributed over the ranges
$-5<\log q<-1$, $-3<\log d<0.3$, in order to identify the parameter
regimes of models that could explain the light curve.
The initial parameters with $\log d>0.3$ were not prepared because of the $d \leftrightarrow d^{-1}$ symmetry.
We find that the only models consistent with the observed light curve
have $q \sim 10^{-4}$ and $d \sim 1$, and that the best fit model has 
$q=(3.95 \pm 0.02) \times 10^{-4}$, $d=0.97537 \pm 0.00007$, 
and other parameters as listed in Table \ref{tb:parameter}. 
The projected position of the planet is pretty close to
the Einstein ring, and therefore $d$ was well constrained.
The light curves and caustic of this event are shown 
in Figure \ref{fig:lightcurve} and Figure \ref{fig:caustic}, respectively, 
which resemble Figure \ref{sec:conclusion} in \citet{Wambsganss1997}.
Here we assumed no orbital motion of the planet around the host star in our model.
So the $d$ and $\alpha$ are the average separation and angle during
half a day when the source is crossing the caustics. The changes of these parameters
due the orbital motion during this period could be same order or slightly larger than
the nominal MCMC error of the average values given above. These changes does not
affect the results in this analysis because they are much smaller than the uncertainty
given in Section \ref{sec:estimate}.

\section{Source Magnitude and Color}
\label{sec:correction}

The dereddened source magnitude and color can be estimated as follows.
We locate the clump in
the color magnitude diagram (CMD) of stars within $2'$ of the
source star, shown in Figure \ref{fig:cmd}, with the following method. 
The stars, which are $I<17$ mag and $(V-I)>1.5$ mag, were used for the clump estimate. 
Among them, the stars within 0.3 mag from the clump centroid were picked up. 
Note that the clump in the first turn was assumed. 
Then, the mean magnitude of $I$ and mean color $(V-I)$ were calculated using the stars within 
0.3 mag and replaced as the new clump centroid. 
This was iterated until the clump centroid position is converged. 
Therefore, we find the clump as $[I, (V-I)]_{\rm clump} = (15.31, 1.91)$. 
The best model source brightness and color are obtained as $[I, (V-I)]_{\rm S} = (19.82, 1.69)$ from the fits. 
With a 0.05 mag correction due to blending by
fainter stars in this crowded field \citep{bennett-ogle109}, this yields
\begin{equation}
[I, (V-I)]_{\rm S} - [I, (V-I)]_{\rm clump} = (4.51, -0.22).
\end{equation}
\noindent
We adopt the dereddened RCG magnitude 
$M_{I,0,\rm clump} = -0.25$ and color $(V-I)_{0,\rm clump} = 1.04$ from 
\citet{Bennett2008}, which is based on 
\citet{Girardi2001} and \citet{Salaris2002}. 
\citet{rat_bar} find that the clump in this field lies 0.12 mag in the
foreground of the Galactic center, which we take to be at 
$R_0 = 8.0 \pm 0.3$ kpc \citep{Yelda2010}. 
Hence, the distance modulus of the clump is DM = 14.40. 
This yields a dereddened RCG centroid in this field of 
\begin{equation}
[I, (V-I)]_{\rm clump, 0} = (14.15, 1.04) \ .
\label{eq:clump}
\end{equation} 
Assuming that the source suffers the same extinction as the clump, we
use the best fit source magnitude and color to obtain the dereddened
values for the source, 
\begin{eqnarray}
[I, (V-I)]_{\rm S, 0} &=& (14.15, 1.04) + (4.51, -0.22) \nonumber \\
				  &=& (18.66, 0.82).
\label{eq-src_colmag}
\end{eqnarray}
A comparison of $(V-I)_{\rm S, 0} $ estimated by this method to 14 
spectra of microlensed source stars at high magnification 
\citep{Bensby2010} suggests
that $(V-I)_{\rm S, 0}$ is determined with an uncertainty of 0.06 mag.
For the uncertainty in $I_{\rm S, 0}$, we estimate uncertainties of 
0.08 from $R_0$, 0.05 from the Galactic bulge RCG centroid, and 0.05 
from the \citet{rat_bar}
offset from the Galactic center, which when added in quadrature yields a total uncertainty of 0.11 mag.

Equation (\ref{eq:clump}) implies extinction of $A_{\rm I} = 1.16 \pm 0.11$ and reddening 
$E(V-I) = 0.87 \pm 0.08$, which is consistent within the error with 
$E(V-I) = 0.97 \pm 0.03$ from the OGLE-II extinction map \citep{Sumi2004}.

\section{\texorpdfstring{Measurement of the Angular Einstein Radius, $\theta_{\rm E}$}{Measurement of the Angular Einstein Radius}}
\label{sec:thetaE}

The sharp caustic crossing features in the MOA-2009-BLG-319 
light curve resolve the finite angular size of the source star, and
these finite source effects allow us to determine the angular 
Einstein radius $\theta_{\rm E}$ and the lens-source relative proper 
motion $\mu_{\rm rel} = \theta_{\rm E}/t_{\rm E}$. Following 
\citet{Yoo2004}, 
we use the dereddened color and magnitude of the source $[I, (V-I)]_{\rm S, 0}$ 
from Eq. (\ref{eq-src_colmag}).
Next, we obtain the source angular radius using the source $V$ and $K$ magnitude. 
We estimate $(V-K)_0$ from $(V-I)_0$ and the \citet{Bessell1988} 
color-color relations for dwarf stars,
\begin{equation}
[K, (V-K)]_{\rm S, 0} = (17.67, 1.81) \pm (0.14, 0.15). 
\label{eq:KandV-K}
\end{equation}
We also estimate the $K$ magnitude using IRSF data, $K_{\rm S,0}=18.09 \pm 0.42$. 
This is consistent with but less accurate than the $K$ magnitude estimated from $(V-I)_0$. 
So, we use $K$ magnitude estimated from $(V-I)_0$.
For main sequence stars, the relationship between color, brightness, 
and a star angular radius $\theta_*$ was determined by 
\citet{Kervella2004} to be
\begin{equation}
\log(2\theta_*) = 0.0755(V-K) + 0.5170 - 0.2K,
\label{eq:source radius}
\end{equation}
which with $K$ and $(V-K)$ from Eq.~(\ref{eq:KandV-K}) implies 
\begin{equation}
\theta_* = 0.66 \pm 0.06 \hspace{0.1cm} \mu \mathrm{as}.
\end{equation}
The fit parameter $\rho \equiv \theta_*/\theta_{\rm E}$ is source star 
angular radius in units of the angular Einstein radius.
Thus, the angular Einstein radius $\theta_{\rm E}$ is
\begin{equation}
\theta_{\rm E} = \frac{\theta_* }{\rho } = 0.34 \pm 0.03 \hspace{0.1cm} \mathrm{mas}.
\end{equation}
Therefore, the source-lens relative proper motion $\mu$ is 
\begin{equation}
\mu = \frac{\theta_{\rm E} }{t_{\rm E} } = 7.52 \pm 0.65 \hspace{0.1cm} \mathrm{mas} \hspace{0.1cm} \mathrm{yr^{-1}}.
\end{equation}

\section{Microlensing Parallax Effect}
\label{sec:parallax}

The event time scale is not long, $t_{\rm E}=16.6$ days, so one does 
not expect to detect the orbital microlensing parallax effect
\citep{refsdal-par,gould-par1,macho-par1}. 
However, the very sharp third peak was observed simultaneously 
from Australia, New Zealand, and Hawaii, i.e., along two nearly 
perpendicular base lines of length, $0.36 R_\oplus $ and 
$1.25 R_\oplus $, respectively. Therefore, there is some chance that
these data will reveal a signal due to terrestrial microlensing parallax 
\citep{hardy95,holz_wald,ogle224}.

Microlensing parallax is usually described by the parallax parameter, 
$\pi_{\rm E}$, which is the amplitude of the two-dimensional microlens parallax 
vector, and the two components of this vector are denoted by
$\pi_{\rm E,\rm E}$ and $\pi_{\rm E,\rm N}$, which are the east and north components of
the vector on the sky.
The microlens parallax vector has the same direction as the lens-source 
proper motion, perpendicular to the line of sight. It is related to the 
lens-source relative parallax $\pi_{\rm rel}$ and the 
angular Einstein radius $\theta_{\rm E}$ \citep{Gould2000} by
\begin{equation}
\pi_{\rm E} = \frac{\pi_{\rm rel}}{\theta_{\rm E}}, \hspace{0.5cm} \pi_{\rm rel} = \pi_{\rm L} - \pi_{\rm S},
\label{eq:microlens_parallax}
\end{equation}
where $\pi_{\rm L}$ and $\pi_{\rm S}$ are the lens and the source parallaxes, respectively.

Our initial search for microlensing parallax included both the orbital and terrestrial
effect, as is necessary for a physically correct model. Our initial fits indicated a weak
microlensing parallax signal, so we searched for orbital parallax and terrestrial parallax
signals separately, in order to determine which type of parallax signal is being
seen and to test for possible systematic errors. We must also consider alternative
model solutions due to the $u_0>0 \leftrightarrow u_0<0$ degeneracy first noted by
\citet{Smith2003}. As the model results listed in Table \ref{tb:parameter} indicate,
orbital parallax can improve the fit $\chi^2$ by only $\Delta\chi^2 = 0.6$, with two
additional parameters, which is not, at all, significant. The best terrestrial parallax 
model, however, does give a formally significant $\chi^2$ improvement
of $\Delta\chi^2 = 6.2$, but this improvement decreases to $\Delta\chi^2 = 6.1$ for
the best physical (terrestrial plus orbital) parallax model. With two additional
parameters, this is formally significant at almost the 
95\% confidence
level. Figure \ref{fig:contour} shows the $\Delta \chi^2$ contours 
for microlensing parallax fits to the MOA-2009-BLG-319 light curve.

The best fit parallax model has $u_0 > 0$ and 
$(\pi_{\rm E,\rm E}, \pi_{\rm E,\rm N}) = (-0.15, 0.15) \pm (0.07, 0.05)$, while the best fit $u_0 < 0$
model has a $\chi^2$ value that is larger than the best fit $u_0 > 0$ solution by
1.7 and only an improvement of $\Delta\chi^2 = 4.4$ over the best fit non-parallax
solution. Thus, the best $u_0 < 0$ model is neither a significant improvement 
over the best non-parallax model nor significantly worse than the best
parallax model. We find that $\chi^2$ improvement for the best fit parallax
model comes from Mt. John
observatory (MOA-II 1.8m and Canterbury 0.6m) telescopes alone, with
a total $\chi^2$ improvement $\Delta\chi^2 = 7.3$, while the contribution
of all the other data sets is $\Delta\chi^2 = -1.2$ (i.e.\ the parallax model is
disfavored). One would expect that $\chi^2$ should improve for the many 
other data sets, and the fact that it does not suggests that the parallax signal
may not be real.

If we assume that the scalar parallax measurement of $\pi_{\rm E}$ is correct,
then it implies that the lens system is located in the inner Galactic disk.
Due to the flat rotation curve of the Galaxy, the stars at this location
are rotating much faster than the typical line of sight to a Galactic bulge
star. As a result, the direction of the parallax vector (which is parallel to
the lens-source relative velocity) is most likely to be in the direction
of Galactic rotation, which is $\sim 30^\circ$ East of North. This is 
similar to the direction of the parallax vector for the best $u_0 < 0$ model, 
but it is roughly perpendicular to that for the $u_0 > 0$ model. So, the 
$u_0 > 0$ solution appears to be disfavored on {\it a priori} grounds.

Because of the low significance of the microlensing parallax signal and the
indications of possible systematic problems with the measurement of the
parallax parameters, we will use only an upper limit on the 
microlensing parallax effect in our analysis.

\section{The Lens Properties}
\label{sec:estimate}

We can place lower limits on the lens mass and distance with our measured angular
Einstein radius, $\theta_{\rm E}$, and our upper limit on the amplitude of the microlens 
parallax vector, $\pi_{\rm E}$. The lens mass is given by
\begin{equation}
M = \frac{\theta_{\rm E}}{\kappa \pi_{\rm E} },
\end{equation}
\noindent
where $\kappa$ = 4$G$/($c^2$ AU) = 8.1439 mas $M_{\odot}^{-1}$. With 
our upper limit from the previous section, $\pi_{\rm E}<0.5$,
gives a lower limit on the total mass of the lens
system, $M>0.08 M_{\odot}$. This implies that the lens primary is
more massive than a brown dwarf and must be a star or stellar remnant.
From Eq. (\ref{eq:microlens_parallax}), this implies that the
source-lens relative parallax is $\pi_{\rm rel} < 0.17\,$mas. 

The vast majority
of source stars for microlensing events seen towards the bulge are stars in 
the bulge, and the MOA-2009-BLG-319 source magnitude and colors are 
consistent with a bulge G-dwarf source. So, it is reasonable to assume that 
the source is a bulge star with a distance of $D_{\rm S} \approx 8.0$ kpc. This
implies that the lens parallax is $\pi_{\rm L} = \pi_{\rm rel} + \pi_{\rm S} < 0.30$ mas, 
from Equation (\ref{eq:microlens_parallax}). The lens parallax is related to the
distance by $\pi_{\rm L} = 1\,{\rm AU}/D_{\rm L}$, so a lower limit on the lens 
distance is $D_{\rm L} > 3.33$ kpc.

An upper limit on the lens mass may be obtained if we assume that the
planetary host star is a main sequence star and not a stellar remnant.
We can consider the blended flux seen at the same location of the source
beyond the measured source flux from the microlensing models.
If we attribute this blended flux to a single star, we can follow
the reasoning of Section \ref{sec:correction} in order estimate 
the dereddened magnitude of the blend star
\begin{equation}
(I, V-I)_{\rm b, 0} = (17.78, 0.75) \pm (0.12, 0.14)\ ,
\end{equation}
under the (conservative) assumption that the blend star lies behind all
the foreground dust. We can now use this as an upper limit on the brightness
of a main sequence lens star. From \citet{Schmidt-Kaler1982} and 
\citet{Bessell1988}, we find an upper limit on the
host star mass of $M < 1.14 M_{\odot }$.

As we found finite source effects in the light curve, we can break out 
one degeneracy of the lens star mass $M$, distance $D_{\rm L}$ and velocity $v$. 
We calculated the probability distribution from Bayesian analysis 
by combining this equation and the measured values of $\theta_{\rm E}$ and $t_{\rm E}$ 
with the Galactic model \citep{Han2003} assuming the distance 
to the Galactic center is 8 kpc. We included the upper limit of microlens parallax amplitude. 
A constraint of the upper limit for blending light was also included 
for the lens mass upper limit. The probability distribution from a Bayesian analysis 
is shown in Figure \ref{fig:likelihood}. The host star is a K or M-dwarf star with a 
mass of $M_{\rm L} = 0.38_{-0.18}^{+0.34}$ $M_\odot$ and distance 
$D_{\rm L} = 6.1_{-1.2}^{+1.1}$ kpc, planetary mass $M_{\rm p} = 50_{-24}^{+44}$ $M_\oplus$ 
and projected separation $r_\perp = 2.0_{-0.4}^{+0.4}$ AU. The physical three-dimensional separation, 
$a = 2.4_{-0.6}^{+1.2}$ AU, was estimated by putting a planetary orbit at random inclination, 
eccentricity and phase \citep{Gould1992}.

\section{Discussion and Conclusion}
\label{sec:conclusion}

We have reported the discovery of a sub-Saturn mass planet in the
light curve of microlensing event, MOA-2009-BLG-319. This event 
was observed by 20 telescopes, the largest number of telescopes to 
participate in a microlensing planet discovery to date. The lens system 
has a mass ratio $q = (3.95 \pm 0.02) \times 10^{-4}$ and a 
separation $d = 0.97537 \pm 0.00007$ Einstein Radii. The lens-source 
relative proper motion was determined to be 
$\mu_{\rm rel} = 7.52 \pm 0.65$ mas~yr$^{-1}$ from the measurement of
finite source effects. A slightly better light curve fit can be obtained when the
(terrestrial) microlensing parallax effect is included in the model, yielding
an improvement of $\Delta \chi^2 = 6.1$. This is very marginal statistical
significance, and there are indications that systematic errors may influence the
result. So, we use our microlensing parallax analysis to set an upper limit
of $\pi_{\rm E}<0.5$. 

The probability distribution estimated from a Bayesian analysis 
indicates that the lens host star mass is $M_{\rm L} = 0.38_{-0.18}^{+0.34}$ $M_\odot$ 
with a sub-Saturn-mass planet, $M_{\rm p} = 50_{-24}^{+44}$ $M_\oplus$ and the 
physical three-dimensional separation $a = 2.4_{-0.6}^{+1.2}$ AU. The 
distance of the lens star is $D_{\rm L} = 6.1_{-1.2}^{+1.1}$ kpc. 
The known microlensing exoplanets are summarized in Figure \ref{fig:exoplanets} 
and Table \ref{tb:exoplanets}.
MOA-2009-BLG-319Lb lies at $\sim 2.3$ times the distance of 
the snow line, which is estimated to be at 
$a_{\mathrm{snow}} = 2.7$ AU~$M/M_\odot$. This is similar to the separation
of other planets found by microlensing \citep{Sumi2010}.

\begin{figure}\begin{center}
\includegraphics[scale=0.3, angle=270, origin=c]{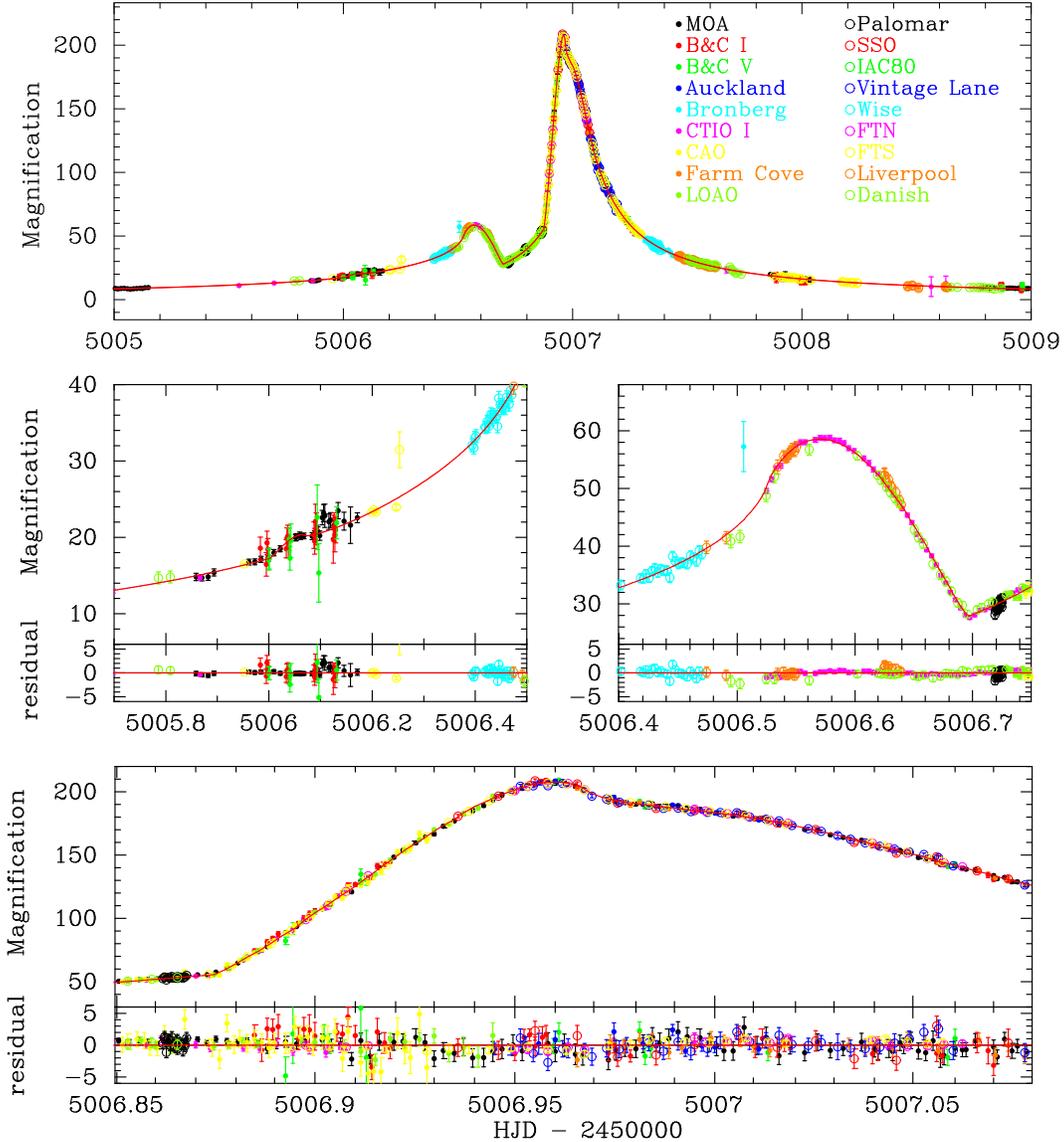} 
\caption{The light curve of planetary microlensing event MOA-2009-BLG-319. 
The top panel shows the data points and the best fit model light curve with finite 
source and limb darkening effects. The three lower panels show close-up views of
the four caustic crossing light curve regions and the residuals from the best fit
light curve. The photometric measurements from MOA, B\&C, Auckland, 
Bronberg, CAO, CTIO, Farm Cove and LOAO are plotted as filled dots
with colors indicated by the legend in the top panel. The other data sets are plotted
with open circles. The data sets of $\mu$FUN Bronberg and SSO 
have been averaged into $0.01\,$day bins, and the RoboNet FTN and FTS 
data sets are shown in $0.005\,$day bins, for clarity.}
\label{fig:lightcurve}
\end{center}\end{figure}

\begin{figure}\begin{center}
\includegraphics[scale=0.25, angle=270, origin=c]{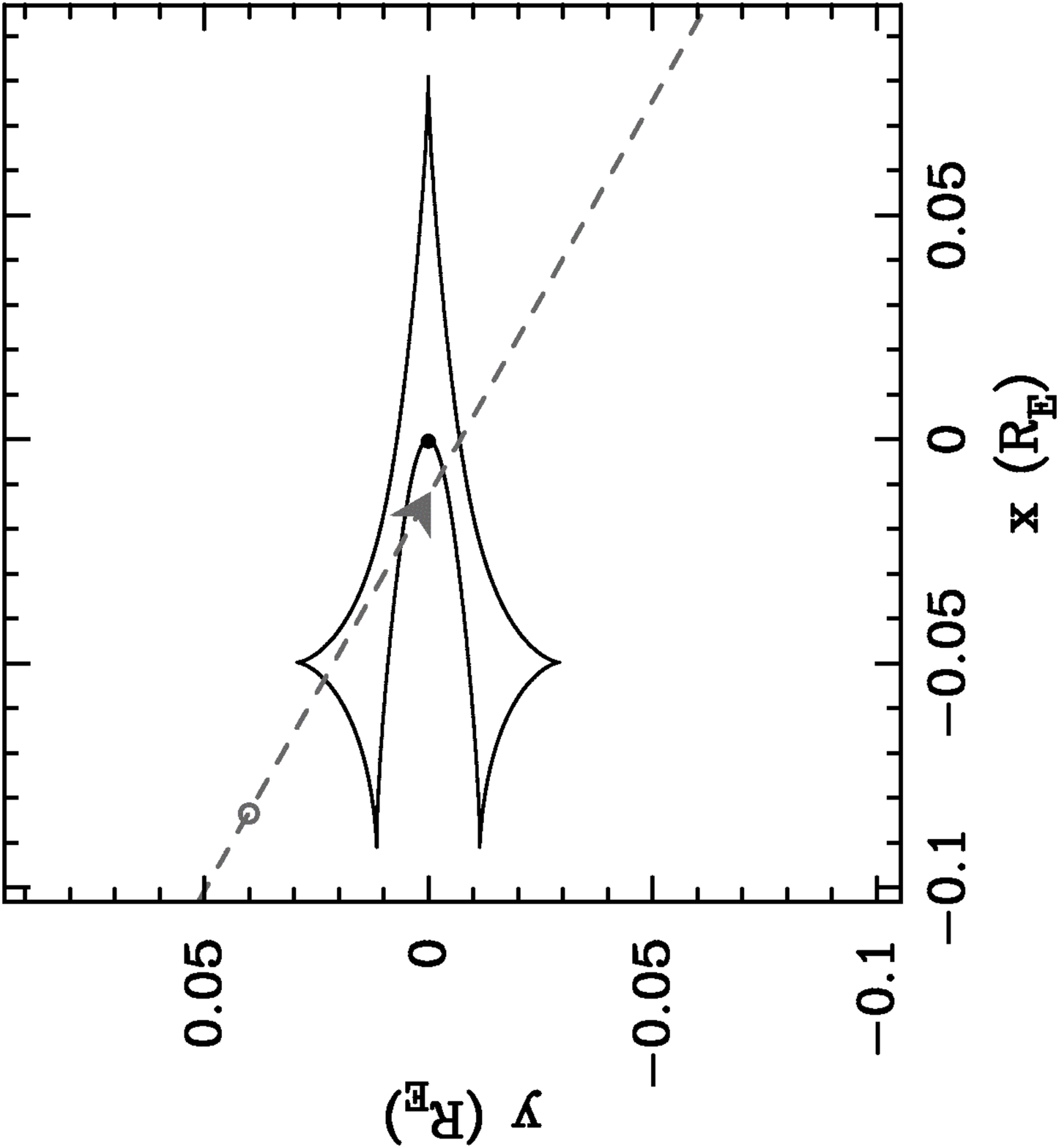} 
\caption{ The caustic is plotted in solid curve for the MOA-2009-BLG-319 
best fit model, and the dash line indicates the source star trajectory. 
The circle represents the source star size. The source star crosses the 
caustic curve four times, with peak magnification of $A_{\rm max} \sim 205$
during the third caustic crossing at ${\rm HJD'} \sim 5006.96$.}
\label{fig:caustic}
\end{center}\end{figure}

\begin{figure}\begin{center}
\includegraphics[scale=0.9, angle=270, origin=c]{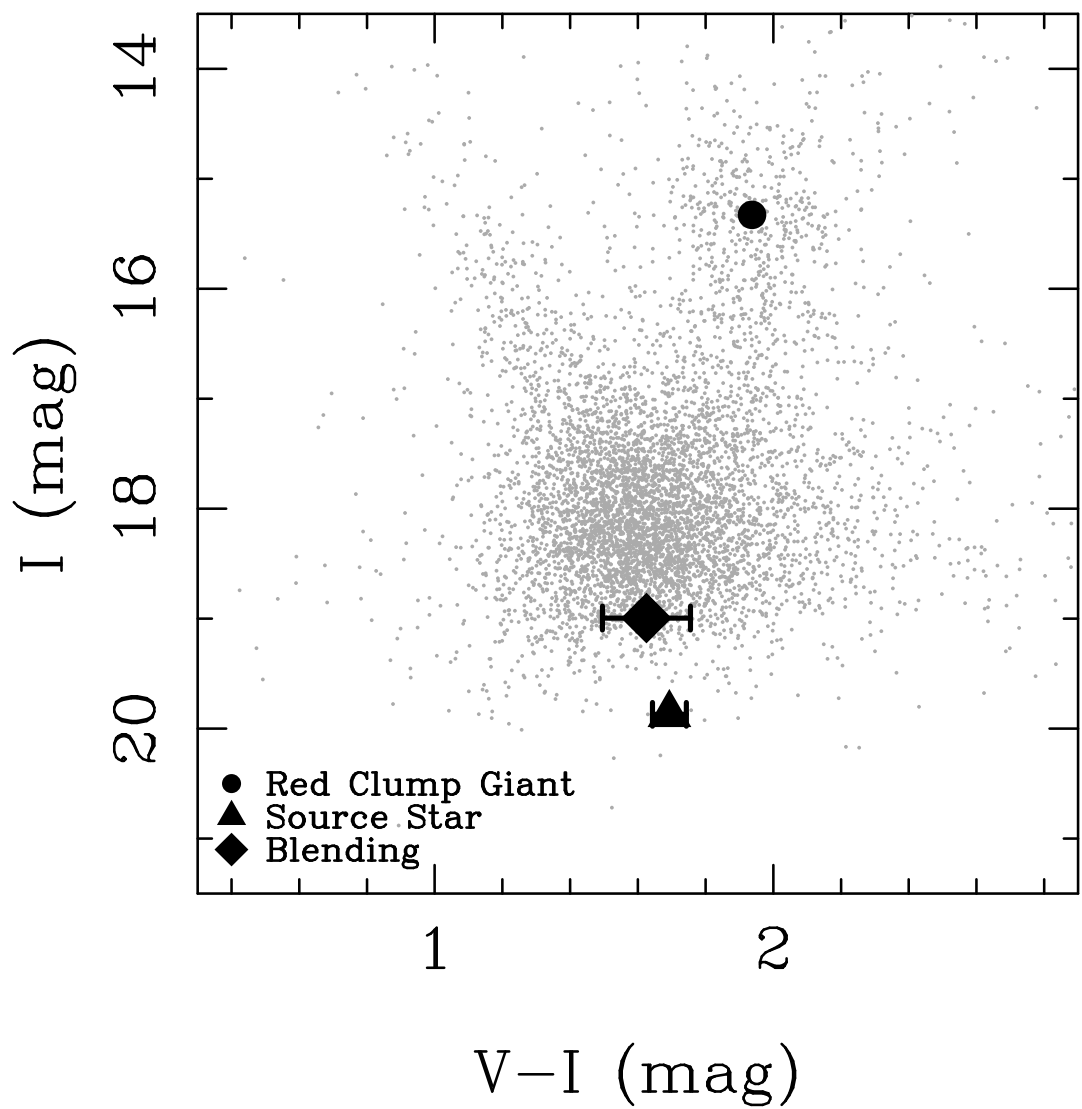} 
\caption{$(V-I, I)$ color magnitude diagram of the stars within
$2'$ of the MOA-2009-BLG-319 source using $\mu$FUN CTIO data
calibrated to OGLE-II. The filled triangle and square indicate the 
source and blend stars, respectively, assuming that the blended light
comes from a single star. The filled circle indicates the center of the
red clump giant distribution.}
\label{fig:cmd}
\end{center}\end{figure}

\begin{figure}
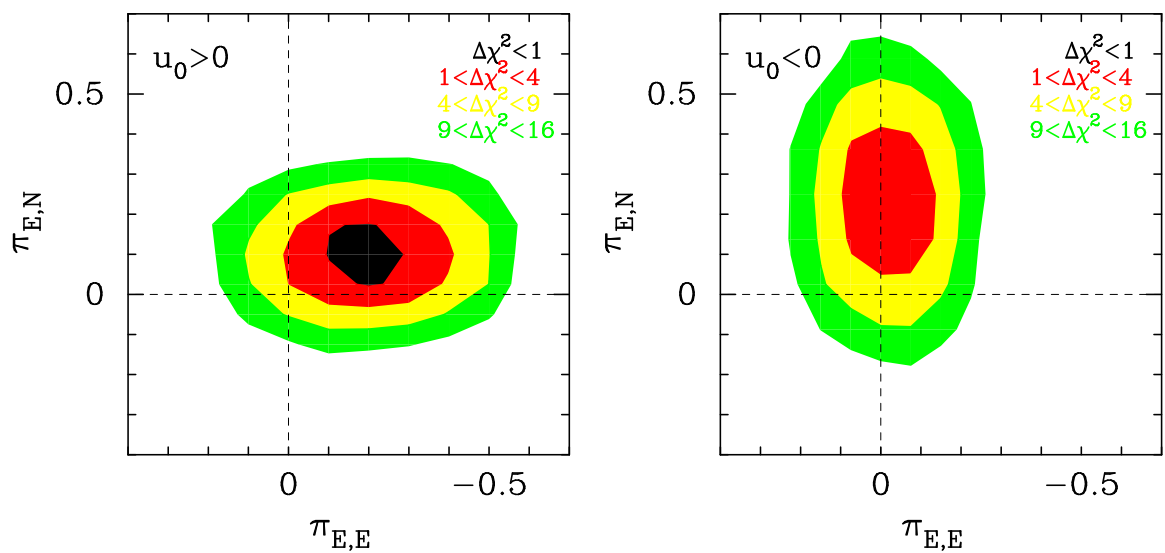
\begin{center}
\begin{tabular}{cc}
\includegraphics[scale=0.55, angle=270, origin=c]{f4a.eps}
&
\includegraphics[scale=0.55, angle=270, origin=c]{f4b.eps}
\end{tabular}
\caption{The contours of $\Delta \chi^2$=1, 4, 9, 16 with orbital and terrestrial 
parallax parameters. The left panel is the result with $u_0>0$ and the right panel 
is with $u_0<0$. The best fit result with $u_0>0$ is better than $u_0<0$ about 
$\Delta \chi^2 = 1.7$. Furthermore, the best fit model with and 
without parallax is different with $\Delta \chi^2 = 6.1$.}
\label{fig:contour}
\end{center}\end{figure}

\begin{figure}\begin{center}
\includegraphics[scale=0.5, angle=270, origin=c]{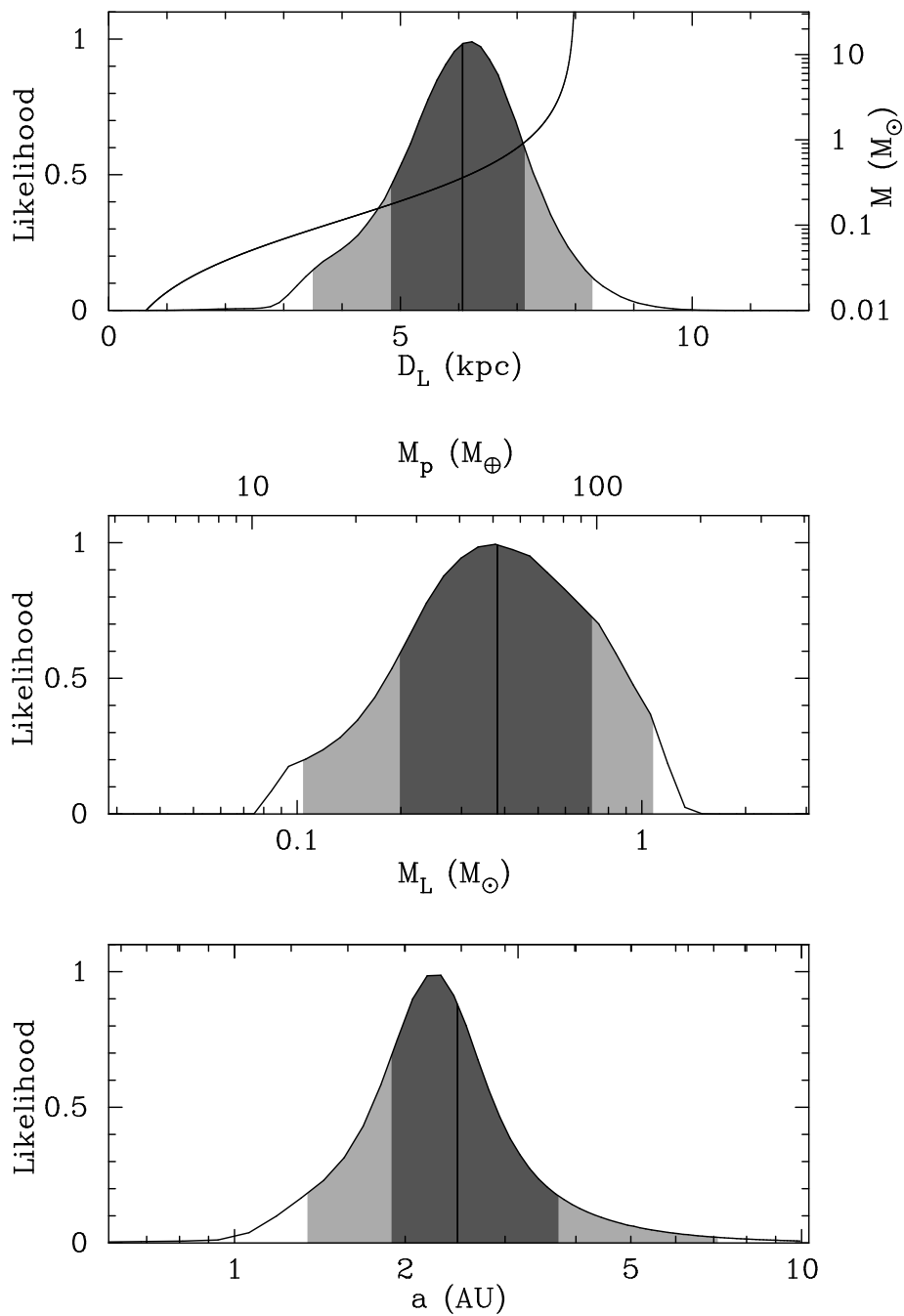}
\caption{Probability distribution from a Bayesian analysis for the distance, 
$D_{\rm L}$, mass, $M_{\rm L}$, and the physical three dimensional separation 
$a$. The vertical solid lines indicate the median values. The dark
and light shaded regions indicate 
the 68\% and 95\%
limits. The solid curve in the top panel indicates the 
mass-distance relation of the lens from the measurement of 
$\theta_{\rm E}$ assuming $D_{\rm S}$ = 8 kpc. Note that $D_{\rm S}$ 
is not fixed in the actual Bayesian analysis.
}
\label{fig:likelihood}
\end{center}\end{figure}

\begin{figure}\begin{center}
\includegraphics[scale=0.3, angle=270, origin=c]{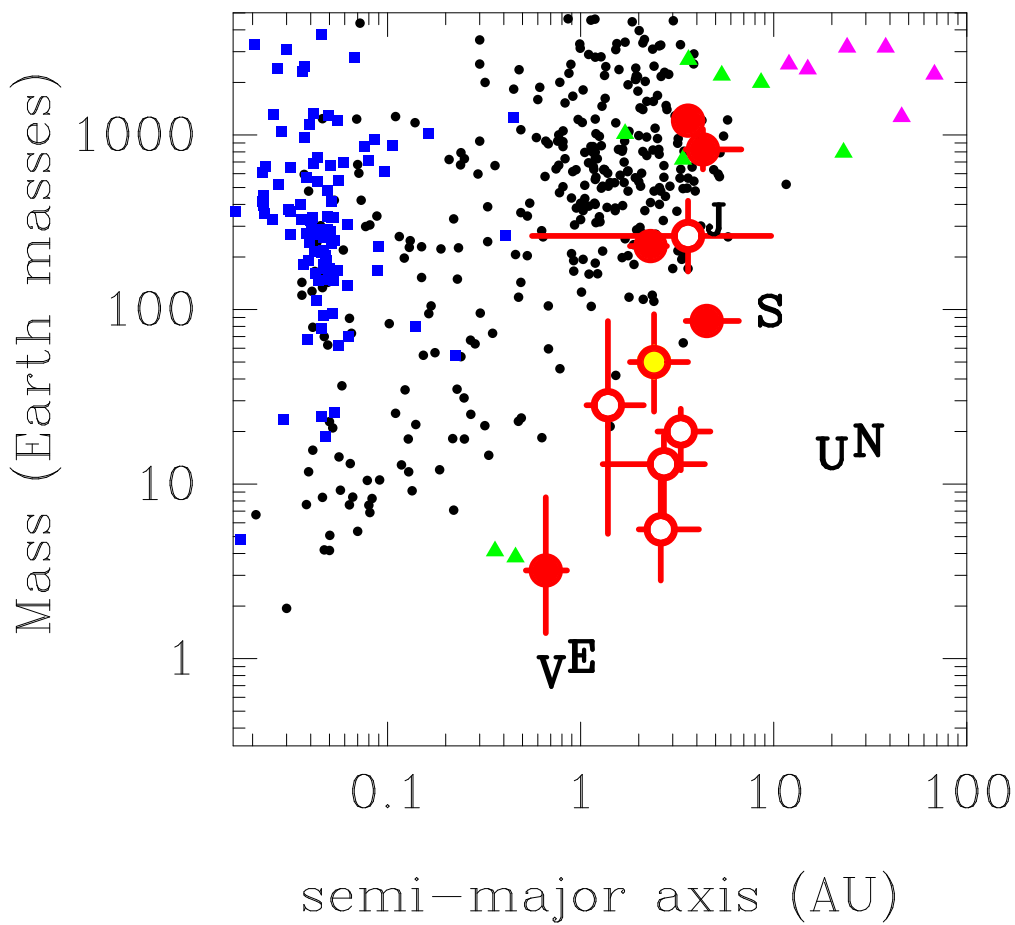}
\caption{
Exoplanets as a function of mass vs. semi-major axis. 
The red circles with error bars indicate planets found by microlensing. 
The filled circles indicate planets with mass measurements, 
while open circles indicate Bayesian mass estimates. 
MOA-2009-BLG-319Lb is indicated by the gold-filled open circle. 
The black dots and blue squares indicate the planets discovered by radial velocities 
and transits, respectively. The magenta and green triangles indicate 
the planets detected via direct imaging and timing, respectively. 
The non-microlensing exoplanet data were taken from The Extrasolar Planets Encyclopaedia 
(\url{http://exoplanet.eu/}). The planets in our solar system are indicated with initial letters.
}
\label{fig:exoplanets}
\end{center}\end{figure}

There is some indications of low level systematic deviations from
the best fit model remaining
in the light curve, near the third and fourth caustic crossing features
(see the bottom panel residuals in Fig.\ref{fig:lightcurve}), 
which does not affect the results in this analysis. 
These systematic light curve deviations might be caused by 
orbital motion of the lens, a second planet, or 
systematic photometry errors. 
A more detailed analysis will be performed in the future 
when the adaptive optics images from the Keck telescope were reduced,
and this analysis may shed more light on the mass
and distance of the host star.

The next few years are expected to see an increase in the rate of 
microlensing planet discoveries. The OGLE group has started 
the OGLE-IV survey with their new $1.4\,{\rm deg}^2$ CCD camera in
March, 2010. This will allow OGLE to survey the bulge at a cadence
almost as high as that of MOA-II, but with better seeing that should yield
a substantial increase in the rate of microlensing planet discoveries.
MOA also plans an upgrade to a $\sim 10\,{\rm deg}^2$ MOA-III CCD camera
in a few years, which will allow an even higher cadence Galactic bulge survey.
The Korean Microlensing Telescope Network (KMTNet) is funded to dramatically
increase the longitude coverage of microlensing survey telescopes. They
plan three wide FOV telescopes to go in South Africa, Australia and South
America. When these telescopes come online, we anticipate another dramatic 
increase in the microlensing planet discovery rate.

\acknowledgements
The MOA project and a part of authors were supported by the Grant-in-Aid for Scientic Research, 
the grant JSPS20340052, JSPS18253002, JSPS Research fellowships, 
the Global COE Program of Nagoya University 
"Quest for Fundamental Principles in the Universe" from JSPS and MEXT of Japan, 
the Marsden Fund of New Zealand, the New Zealand Foundation for Research and Technology, 
and grants-in-aid from Massey University and the University of Auckland.
N.M. was supported by JSPS Research Fellowships for Young Scientists. 
T.S. was supported by the grant JSPS20740104. 
D.P.B. was supported by grants NNX07AL71G and NNX10AI81G from NASA and
AST-0708890 from the NSF.
A.G. and S.D. were supported in part by NSF AST-0757888.
A.G., S.D., S.G., and R.P. were supported in part by NASA NNG04GL51G.  
Work by S.D. was performed in part
under contract with the California Institute of Technology (Caltech)
funded by NASA through the Sagan Fellowship Program.
J.C.Y. was supported by a NSF Graduate Research Fellowship.
Work of C.H. was supported by the Creative Research Initiative
program (2009-0081561) of the National Research Foundation of Korea. 
Astronomical research at Armagh Observatory was supported by the
Department of Culture, Arts and Leisure (DCAL), Northern Ireland, UK.
F.F., D.R. and J.S. acknowledge support from the Communaut\'{e} fran\c{c}aise de Belgique 
-- Actions de recherche concert\'{e}es -- Acad\'{e}mie universitaire Wallonie-Europe

\clearpage


\clearpage


\begin{deluxetable}{ccccccc}
\tablecaption{Limb darkening coefficients for the source star with effective temperature $T_{\rm eff}$=5500~K, surface gravity $\log g$=4.5 and metallicity $\log$[M/H]=0.0 \citep{Claret2000}.
\label{tb:LDparameter}
}
\tablewidth{0pt}
\tablehead{
\colhead{filter	color} & \colhead{$V$} & \colhead{$R$} & \colhead{$I$} & \colhead{$J$} & \colhead{$H$} & \colhead{$K$} 
}

\startdata
$c$				&	0.3866	&	0.2556	&	0.1517	&	-0.0234	&	-0.2154	&	-0.1606	\\
$d$				&	0.4303	&	0.5027	&	0.5281	&	 0.6021	&	 0.7695	&	 0.6324	\\
\enddata

\end{deluxetable}

\begin{deluxetable}{cccc|crrrrrrrrrr}
\tabletypesize{\scriptsize}
\rotate
\tablecaption{The best fit model parameters with various effects, finite source, orbital and terrestrial parallax, and $u_0$. The lines with "$\sigma $" list the 1$\sigma $ error of parameters given by MCMC. HJD'$\equiv $HJD-2450000. Note that the $u_0$ conventions are the same as Fig. 2 of \citet{Gould2004}. $\chi^2 $ value is the result of the fitting with 18 data sets, which have 7210 data points. The model search with finite source and orbital parallax effects were done by a grid search.}
\tablewidth{0pt}
\tablehead{
\colhead{orbital}     & \colhead{terrestrial} & \colhead{$u_0>0$} & \colhead{$u_0<0$} & & 
\colhead{$t_0$} & \colhead{$t_{\rm E}$}     & \colhead{$u_0$}     & \colhead{$q$}    & \colhead{$d$}    & \colhead{$\alpha$} & 
\colhead{$\rho$}  & \colhead{$\pi_{\rm E,\rm E}$} & \colhead{$\pi_{\rm E,\rm N}$} & \colhead{$\chi^2$} \\

\colhead{parallax}    & \colhead{parallax}    & & & & 
\colhead{HJD'}    & \colhead{[days]}      & \colhead{$10^{-3}$}   & \colhead{$10^{-4}$} &               & \colhead{[rad]}    & 
\colhead{$10^{-3}$} &  &  &          

\label{tb:parameter}
}

\startdata
         &         & $\circ$ &         & &
5006.99482 & 16.57 & 6.22 & 3.95 & 0.97537 & 5.7677 & 1.929 & \nodata & \nodata & 7023.8 \\ 

         &         &         &         & $\sigma$ &
0.00006 & 0.08 & 0.03 & 0.02 & 0.00007 & 0.0005 & 0.010 & \nodata & \nodata &     \\ 

         &         &         & $\circ$ &  &
5006.99485 & 16.56 & -6.23 & 3.95 & 0.97540 & 0.5156 & 1.931 & \nodata & \nodata & 7023.8 \\ 

         &         &         &         & $\sigma$ &
0.00005 & 0.08 & 0.03 & 0.02 & 0.00006 & 0.0005 & 0.009 & \nodata & \nodata &     \\ 

 $\circ$ &         & $\circ$ &         & &
5006.99480 & 16.59 & 6.22 & 3.95 & 0.97540 & 5.7673 & 1.929 & 0.40 & 0.30 & 7023.2 \\ 

         &         &         &         & $\sigma$ &
0.00007 & 0.09 & 0.03 & 0.02 & 0.00007 & 0.0005 & 0.011 & \nodata & \nodata &     \\ 

 $\circ$ &         &         & $\circ$ & &
5006.99482 & 16.56 & -6.23 & 3.95 & 0.97534 & 0.5155 & 1.931 & 0.40 & -0.30 & 7023.4 \\ 

         &         &         &         & $\sigma$ &
0.00006 & 0.08 & 0.03 & 0.02 & 0.00007 & 0.0005 & 0.010 & \nodata & \nodata &     \\ 

         & $\circ$ & $\circ$ &         & &
5006.99477 & 16.61 & 6.21 & 3.94 & 0.97540 & 5.7671 & 1.926 & -0.23 & 0.12 & 7017.6 \\ 

         &         &         &         & $\sigma$ &
0.00006 & 0.08 & 0.03 & 0.02 & 0.00007 & 0.0005 & 0.010 & 0.07 & 0.04 &     \\ 

         & $\circ$ &         & $\circ$ & &
5006.99483 & 16.57 & -6.23 & 3.95 & 0.97542 & 0.5161 & 1.931 & -0.02 & 0.26 & 7019.2 \\ 

         &         &         &         & $\sigma$ &
0.00006 & 0.07 & 0.03 & 0.02 & 0.00007 & 0.0005 & 0.010 & 0.04 & 0.07 &     \\ 

 $\circ$ & $\circ$ & $\circ$ &         & &
5006.99478 & 16.60 & 6.21 & 3.94 & 0.97540 & 5.7673 & 1.926 & -0.15 & 0.15 & 7017.7 \\ 

         &         &         &         & $\sigma$ &
0.00006 & 0.08 & 0.03 & 0.02 & 0.00007 & 0.0004 & 0.010 & 0.07 & 0.05 &     \\ 

 $\circ$ & $\circ$ &         & $\circ$ & &
5006.99481 & 16.56 & -6.23 & 3.95 & 0.97538 & 0.5162 & 1.932 & -0.04 & 0.23 & 7019.4 \\ 

         &         &         &         & $\sigma$ &
0.00006 & 0.07 & 0.03 & 0.02 & 0.00007 & 0.0005 & 0.009 & 0.04 & 0.07 &     \\ 

\enddata

\end{deluxetable}

\begin{deluxetable}{rccrcl}
\tabletypesize{\footnotesize}
\rotate
\tablecaption{
Parameters of exoplanets discovered by microlensing. MOA-2007-BLG-400Lb has two solutions due to a strong close/wide model degeneracy, and details of the MOA-2008-BLG-310Lb parameters are discussed by \citet{moa310} and \citet{Sumi2010}.
\label{tb:exoplanets}
}
\tablewidth{0pt}
\tablehead{
\colhead{name} & \colhead{Host Star Mass}   & \colhead{Distance} & \colhead{Planet Mass}  & \colhead{Separation} & \colhead{Mass estimated by}\\
\colhead{}     & \colhead{$M_{\rm L}(M_\odot)$} & \colhead{$D_{\rm L}$(kpc)} & \colhead{$M_{\rm p}$} & \colhead{$a$(AU)} & \colhead{} 
}

\startdata
OGLE-2003-BLG-235Lb & $ 0.63^{\ +0.07}_{\ -0.09}$ & $ 5.8^{\ +0.6}_{\ -0.7}$ & $ 2.6^{\ +0.8}_{\ -0.6} \ M_{\rm J}$ & $ 4.3^{\ +2.5}_{\ -0.8}$ & $\theta_{\rm E}$, lens brightness \\ 
OGLE-2005-BLG-071Lb & $ 0.46 \pm 0.04$ & $ 3.2 \pm 0.4$ & $ 3.8 \pm 0.4 \ M_{\rm J}$ & $ 3.6 \pm 0.2$ & $\theta_{\rm E}, \pi_{\rm E}$, detection of the lens\\ 
OGLE-2005-BLG-169Lb & $ 0.49^{\ +0.23}_{\ -0.29}$ & $ 2.7^{\ +1.6}_{\ -1.3}$ & $ 13^{\ +6}_{\ -8} \ M_\oplus$ & $ 2.7^{\ +1.7}_{\ -1.4}$ & $\theta_{\rm E}$, Bayesian \\ 
OGLE-2005-BLG-390Lb & $ 0.22^{\ +0.21}_{\ -0.11}$ & $ 6.6^{\ +1.0}_{\ -1.0}$ & $ 5.5^{\ +5.5}_{\ -2.7} \ M_\oplus$ & $ 2.6^{\ +1.5}_{\ -0.6}$ & $\theta_{\rm E}$, Bayesian\\ 
OGLE-2006-BLG-109Lb & $ 0.51^{\ +0.05}_{\ -0.04}$ & $ 1.49 \pm 0.19$ & $ 231 \pm 19 \ M_\oplus$ & $ 2.3 \pm 0.5$ & $\theta_{\rm E}, \pi_{\rm E}$\\ 
                  c & $ ^{}_{}$ & $ ^{}_{}$ & $ 86 \pm 7 \ M_\oplus$ & $ 4.5^{\ +2.1}_{\ -1.0}$ & $\theta_{\rm E}, \pi_{\rm E}$\\
OGLE-2007-BLG-368Lb & $ 0.64^{\ +0.21}_{\ -0.26}$ & $ 5.9^{\ +0.9}_{\ -1.4}$ & $ 20^{\ +7}_{\ -8} \ M_\oplus$ & $ 3.3^{\ +1.4}_{\ -0.8}$ & $\theta_{\rm E}$, Bayesian\\ 
MOA-2007-BLG-192Lb & $ 0.084^{\ +0.015}_{\ -0.012}$ & $ 0.70^{\ +0.21}_{\ -0.12}$ & $ 3.2^{\ +5.2}_{\ -1.8} \ M_\oplus$ & $ 0.66^{\ +0.19}_{\ -0.14}$ & $\theta_{\rm E}, \pi_{\rm E}$\\ 
MOA-2007-BLG-400Lb & $ 0.30^{\ +0.19}_{\ -0.12}$ & $ 5.8^{\ +0.6}_{\ -0.8}$ & $ 0.83^{\ +0.49}_{\ -0.31} \ M_{\rm J}$ & $ 0.72^{\ +0.38}_{\ -0.16}\ /\ 6.5^{\ +3.2}_{\ -1.2}$ & $\theta_{\rm E}$, Bayesian\\ 
MOA-2008-BLG-310Lb & $ 0.67 \pm 0.14$ & $ >6.0$ & $ 28 ^{\ +58}_{\ -23} \ M_\oplus$ & $ 1.4 ^{\ +0.7}_{\ -0.3}$ & $\theta_{\rm E}$, Bayesian\\ 
MOA-2009-BLG-319Lb & $ 0.38^{\ +0.34}_{\ -0.18}$ & $ 6.1^{\ +1.1}_{\ -1.2}$ & $ 50^{\ +44}_{\ -24} \ M_\oplus$ & $ 2.4^{\ +1.2}_{\ -0.6}$ & $\theta_{\rm E}$, Bayesian\\ 
\enddata

\end{deluxetable}


\begin{thebibliography}{}

\bibitem[Albrow et al.(2009)]{Albrow2009} 
   Albrow,~M.~D., et al. 2009, \mnras, 397, 2099   

\bibitem[Alcock et al.(1995)]{macho-par1}
   Alcock,~C., et al. 1995, \apj, 454, L125
   
\bibitem[Beaulieu et al.(2006)]{ogle390}
   Beaulieu,~J.-P., et al.\ 2006, \nat, 439, 437
   
\bibitem[Bennett(2010)]{Bennett2010} 
   Bennett,~D.~P. 2010, \apj, 716, 1408 

\bibitem[Bennett \& Rhie(1996)]{Bennett1996}
   Bennett,~D.~P., \& Rhie,~S.~H. 1996, \apj, 472, 660
   
\bibitem[Bennett et al.(2008)]{Bennett2008} 
   Bennett,~D.~P., et al. 2008, \apj, 684, 663

\bibitem[Bennett et al.(2010)]{bennett-ogle109}
   Bennett,~D.~P., et al. 2010, \apj, 713, 837

\bibitem[Bensby et al.(2010)]{Bensby2010} 
   Bensby,~T., et al. 2010, \aap, 512, A41  

\bibitem[Bessell \& Brett(1988)]{Bessell1988} 
   Bessell,~M.~S., \& Brett,~J.~M. 1988, \pasp, 100, 1134  

\bibitem[Bond et al.(2001)]{Bond2001} 
   Bond,~I.~A., et al. 2001, \mnras, 327, 868  

\bibitem[Bond et al.(2004)]{Bond2004} 
   Bond,~I.~A., et al. 2004, \apj, 606, L155 

\bibitem[Bramich(2008)]{Bramich2008} 
   Bramich,~D.~M. 2008, \mnras, 386, L77 

\bibitem[Claret(2000)]{Claret2000} 
   Claret, A. 2000, \aap, 363, 1081  

\bibitem[Cumming et al.(2008)]{Cumming2008} 
   Cumming,~A., et al. 2008, \pasp, 120, 531  

\bibitem[Dong et al.(2009)]{dong-moa400} 
   Dong,~S., et al.\ 2009, \apj, 698, 1826

\bibitem[Doran \& Mueller(2004)]{DoranMueller2004} 
   Doran,~M., \& Mueller,~C.~M. 2004, \jcap, 9, 3

\bibitem[Gaudi et al.(2008)]{Gaudi2008} 
   Gaudi,~B.~S., et al. 2008, Science, 319, 927 

\bibitem[Girardi \& Salaris(2001)]{Girardi2001} 
   Girardi,~L., \& Salaris,~M. 2001, \mnras, 323, 109  

\bibitem[Gould(1992)]{gould-par1}
   Gould,~A.\ 1992, \apj, 392, 442
    
\bibitem[Gould(2000)]{Gould2000} 
   Gould,~A. 2000, \apj, 542, 785   

\bibitem[Gould(2004)]{Gould2004} 
   Gould,~A. 2004, \apj, 606, 319  

\bibitem[Gould \& Loeb(1992)]{Gould1992} 
   Gould,~A., \& Loeb,~A. 1992, \apj, 396, 104  

\bibitem[Gould et al.(2006)]{Gould2006} 
   Gould,~A., et al. 2006, \apj, 644, L37  

\bibitem[Gould et al.(2009)]{ogle224}
   Gould,~A., et al.\ 2009, \apjl, 698, L147 
   
\bibitem[Gould et al.(2010)]{Gould2010} 
   Gould,~A., et al. 2010, \apj, 720, 1073  

\bibitem[Griest \& Safizadeh(1998)]{griest_saf} 
   Griest,~K., \& Safizadeh,~N.\ 1998, \apj, 500, 37

\bibitem[Han \& Gould(2003)]{Han2003} 
   Han,~C., \& Gould,~A. 2003, \apj, 592, 172 

\bibitem[Hardy \& Walker(1995)]{hardy95}
   Hardy,~S.~J., \& Walker,~M.~A. 1995, \mnras, 276, L79
   
\bibitem[Holz \& Wald(1996)]{holz_wald}
   Holz,~D.~E., \& Wald,~R.~M.\ 1996, \apj, 471, 64
    
\bibitem[Ida \& Lin(2004)]{ida_lin}
   Ida,~S., \& Lin,~D.~N.~C.\ 2004, \apj, 616, 567
    
\bibitem[Janczak et al.(2010)]{moa310}
   Janczak,~J., et al.\ 2010, \apj, 711, 731
   
\bibitem[Kennedy \& Kenyon(2008)]{kennedy_snowline}
   Kennedy,~G.~M., \& Kenyon,~S.~J.\ 2008, \apj, 673, 502 
   
\bibitem[Kervella et al.(2004)]{Kervella2004} 
   Kervella,~P., et al. 2004, \aap, 426, 297 

\bibitem[Lecar et al.(2006)]{lecar_snowline}
   Lecar,~M., Podolak,~M., Sasselov,~D., \& Chiang,~E.\ 2006, \apj, 640, 1115
   
\bibitem[Rattenbury et al.(2002)]{Rattenbury2002} 
   Rattenbury,~N.~J., et al. 2002, \mnras, 335, 159

\bibitem[Rattenbury et al.(2007)]{rat_bar}
   Rattenbury,~N.~J., Mao,~S., Sumi,~T., \& Smith,~M.~C.\ 2007, \mnras, 378, 1064

\bibitem[Refsdal(1966)]{refsdal-par}
   Refsdal,~S. 1966, \mnras, 134, 315

\bibitem[Rhie et al.(2000)]{mps-98blg35}
   Rhie,~S.~H.,~et al.\ 2000, \apj, 533, 378

\bibitem[Sako et al.(2008)]{Sako2008} 
   Sako,~T., et al. 2008, Exp. Astron., 22, 51 

\bibitem[Salaris \& Girardi(2002)]{Salaris2002} 
   Salaris,~M., \& Girardi,~L. 2002, \mnras, 337, 332 

\bibitem[Schmidt-Kaler(1982)]{Schmidt-Kaler1982}  
   Schmidt-Kaler,~Th. 1982, in Landolt-B\"ornstein: Numerical Data and Functional Relationships in Science and Technology, Vol 2b, ed. K. Schaifers \& H. H. Voigt (Berlin: Springer)  

\bibitem[Smith et al.(2003)]{Smith2003} 
   Smith,~M.~C., Mao,~S., \& Paczy\'nski,~B. 2003, \mnras, 339, 925  

\bibitem[Sumi(2004)]{Sumi2004} 
   Sumi,~T. 2004, \mnras, 349, 193 

\bibitem[Sumi et al.(2003)]{Sumi2003} 
   Sumi,~T., et al. 2003, \apj, 591, 204 

\bibitem[Sumi et al.(2010)]{Sumi2010} 
   Sumi,~T., et al. 2010, \apj, 710, 1641 

\bibitem[Udalski(2003)]{Udalski2003} 
   Udalski,~A. 2003, \actaa, 53, 291 

\bibitem[Udalski et al.(2005)]{ogle71}
   Udalski,~A., et al.\ 2005, \apjl, 628, L109
   
\bibitem[Verde et al.(2003)]{Verde2003} 
   Verde,~L., et al. 2003, \apjs, 148, 195 

\bibitem[Wambsganss(1997)]{Wambsganss1997} 
   Wambsganss,~J.\ 1997, \mnras, 284, 172 

\bibitem[Wozniak(2000)]{Wozniak2000} 
   Wozniak,~P.~R. 2000, \actaa, 50, 421 

\bibitem[Yelda et al.(2010)]{Yelda2010} 
   Yelda,~S., et al. 2010, arXiv1002.1729 
 
\bibitem[Yoo et al.(2004)]{Yoo2004} 
   Yoo,~J., et al. 2004, \apj, 603, 139 

\end{thebibliography}
\end{document}